\documentclass{jetpl}
\twocolumn

\usepackage{amsmath,amssymb,epsfig,color,pstricks,bm,graphics,alltt}

\usepackage[colorlinks,plainpages=false,linkcolor=black,urlcolor=blue,citecolor=black,pdfpagemode=UseNone,pdfstartview=FitBH]{hyperref}

\definecolor{MyDarkBlue}{rgb}{0,  0.3,  0.9}
\definecolor{MyDarkBlack}{rgb}{0,  0,  0}
\newcommand \modified[1]{\textcolor{black}{#1}}

\begin{document}

%%% article in English
\lat

%%% article title
\title{LDA$^\prime$+DMFT Investigation of Electronic Structure of K$_{1-x}$Fe$_{2-y}$Se$_2$
Superconductor}

%%% article title - for colontitle (at the top of the page)
\rtitle{LDA$^\prime$+DMFT Investigation of hole doped K$_{1-x}$Fe$_{2-y}$Se$_2$}

%%% article title - for table of contents (usualy identical with \title)
\sodtitle{LDA$^\prime$+DMFT Investigation of Electronic Structure of K$_{1-x}$Fe$_{2-y}$Se$_2$ Superconductor}

%%% author(s) ( + e-mail)
\author{$^{a}$I.\ A.\ Nekrasov\thanks{E-mail: nekrasov@iep.uran.ru},
$^{a}$N.\ S.\ Pavlov\thanks{E-mail: pavlov@iep.uran.ru}, 
$^{a,b}$M.\ V.\ Sadovskii\thanks{E-mail: sadovski@iep.uran.ru}}

%%% author(s) - for colontitle (at the top of the page)
\rauthor{I.\ A.\ Nekrasov, N.\ S.\ Pavlov, M.\ V.\ Sadovskii}

%%% author(s) - for table of contents
\sodauthor{Nekrasov, Pavlov, Sadovskii}

%%% author(s) - for table of contents
\sodauthor{Nekrasov, Pavlov, Sadovskii}

%%% author's address(es)
\address{$^a$Institute for Electrophysics, Russian Academy of Sciences, 
Ural Branch, Amundsen str. 106,  Ekaterinburg, 620016, Russia\\
$^b$Institute for Metal Physics, Russian Academy of Sciences, Ural Branch,
S.Kovalevskoi str. 18, Ekaterinburg, 620990, Russia}

%%% dates of submition & resubmition (if submitted once, second argument is *)
\dates{November 2012}{*}

%%% abstract
\abstract{
We investigate electronic structure of the new iron chalcogenide
high temperature superconductor K$_{1-x}$Fe$_{2-y}$Se$_2$ (hole doped
case with $x=0.24$, $y=0.28$) in the normal phase using the novel 
LDA$^\prime$+DMFT computational approach. 
We show that this iron chalcogenide is more correlated
in a sense of bandwidth renormalization (\modified{energy scale compression by factor about 5 in the interval $\pm$1.5~eV)}, than 
typical iron pnictides (compression factor about 2), though the Coulomb interaction 
strength is almost the same in both families.
Our results for spectral densities are in general agreement with
recent ARPES data on this system. It is found that all Fe-3d($t_{2g}$) 
bands crossing the Fermi level have equal renormalization, in contrast to 
some previous interpretations. Electronic states at the Fermi level 
are of predominantly $xy$ symmetry. Also we show that LDA$^\prime$+DMFT 
results are in better agreement with experimental spectral function maps, 
than the results of conventional LDA+DMFT. Finally we make predictions for
photoemission spectra lineshape for K$_{0.76}$Fe$_{1.72}$Se$_2$.
}


\maketitle

The iron based FeAs(Se) high-temperature superconductors \cite{kamihara_08}
are one of the hottest topics of the present day condensed matter research
\cite{UFN_90,Hoso_09,FeSe}.
Recent discovery of iron chalcogenides K$_x$Fe$_2$Se$_2$ \cite{Guo10}, 
Cs$_x$Fe$_2$Se$_2$ \cite{Krzton10} and (Tl,K)Fe$_x$Se$_2$ \cite{Fang} 
with superconducting critical temperatures T$_c$ around 30K, which is typical 
for the most studied 122 iron pnictides \cite{rott,ChenLi,Chu,Bud} 
initiated intensive studies of these new systems, which was further stimulated
by the discovery of nontrivial antiferromagnetic 
ordering with very high Neel temperature about 550K and Fe vacancies ordering 
at approximately the same temperatures in K$_{0.8}$Fe$_{1.6}$Se$_2$
(the so called 245 phase) \cite{Fe_order}. Because of the complicated 
picture of microscopic phase separation in this system, there is still no 
consensus on the composition of the phase, responsible for superconductivity,
though the majority point of view indicate to KFe$_2$Se$_2$ (122 phase) as a
parent compound for superconductivity (while 245 phase is insulating) 
\cite{MISM,Kord,Wen}. There is also experimental evidence for some other phases 
being present in this system \cite{WLi}. 

From crystal structure point of view AFe$_2$As$_2$, Fe(Se,Te) and AFe$_2$Se$_2$ systems
are formed by identical layers of Fe(As,Se)$_4$ tetrahedra.
AFe$_2$As$_2$ and AFe$_2$Se$_2$ compounds are isostructural.
LDA (local density approximation) calculated electronic band structures of
Fe(Se,Te) \cite{SinghFeSe} and AFe$_2$As$_2$ \cite{Nekr2,Shein,Krell}
are quite similar to each other, especially if we are dealing only with 
bands in the vicinity of the Fermi level. LDA electronic structure of 
AFe$_2$Se$_2$ is significantly different \cite{Shein_kfese,Nekr_kfese}.
Direct comparison of LDA band structures of Fe pnictides and chalcogenides
can be found in Refs.~\cite{Nekr_kfese,MISM}.

From the early days of iron based superconductors, it was noted that electronic 
correlations are important for understanding the physics of pnictide materials 
\cite{Haule,Craco,Shorikov,Ba122_DMFT}. Electronic correlations for these materials were 
taken into account within LDA+DMFT hybrid computational scheme \cite{LDADMFT}.
It is rather common opinion now, that the main effect of correlations onto 
band structure of Fe pnictides reduces to the simple LDA bandwidth
renormalization (narrowing) by the factor of 2 or 3. There are only few papers 
on LDA+DMFT in Fe chalcogenides up to now \cite{DMFT_FeSe,DMFT_AFeSe}.

The AFe$_2$Se$_2$ systems were rather extensively studied by angular resolved
photoemission spectroscopy (ARPES) \cite{ARPES_AFeSe}. In contrast to 
AFe$_2$As$_2$ compounds, with several, more or less well defined, hole cylinders
of the Fermi surface around $\Gamma$-point, ARPES data for AFe$_2$Se$_2$ show
rather weak indications for Fermi surface around $\Gamma$-point. Around 
($\pi,\pi$) point in both classes of superconductors electron Fermi surface 
sheets are well observed. These ARPES results are in rough agreement with
LDA predictions \cite{Shein_kfese,Nekr_kfese,MISM}.

This paper was inspired by the recent ARPES work \cite{exp} on 
A$_x$Fe$_{2-y}$Se$_2$ (A=K,Rb). Here we present our LDA+DMFT and 
LDA$^\prime$+DMFT \cite{CLDA} results for hole doped 
K$_{0.76}$Fe$_{1.72}$Se$_2$. 
LDA and LDA$^\prime$ calculations were performed using Linearized 
Muffin-Tin Orbitals method \cite{LMTO}, with settings
described in Ref.~\cite{Nekr_kfese}. To solve DMFT effective five orbital
impurity problem we used the Hirsh-Fye Quantum Monte-Carlo algorithm \cite{QMC},
with temperature about 280K. LDA+DMFT and LDA$^\prime$+DMFT
densities of states and spectral functions were obtained as proposed in
Ref.~\cite{CLDA}. Coulomb parameter $U$ was taken to be 3.75~eV and Hund
parameter -- $J$=0.7~eV \cite{exp}. These parameters agree well with those
calculated in Ref.~\cite{param}. To define DMFT lattice problem we used the
full (i.e. without any downfolding or projecting) LDA Hamiltonian, which 
included Fe-3d, Se-4p and K-4s states.

In Fig.~1 we plot LDA (dashed lines) and LDA$^\prime$ (solid lines) calculated 
bands dispersions (right panel), as well as total, Fe-3d and Se-4p densities of 
states (left panel) for stoichiometric KFe$_2$Se$_2$. The Se-4p states are 
located between -7eV and -3.5eV. The Se-4p states are well separated in energy 
from Fe-3d states which cross the Fermi level. The Fe-3d states expand from 
-2.2 eV up to +1 eV. This is the same or similar to previous LDA results of 
Refs.~\cite{Shein_kfese,Nekr_kfese}. For LDA$^\prime$ results we observe 
band shapes, that are almost identical to those of LDA, with approximately 
rigid shift of Se-4p states down in energy for LDA$^\prime$ \cite{CLDA}.
Note the effect of LDA$^\prime$ -- splitting of $xy$ bands around 0.55 eV.

\begin{figure}[!t]
\includegraphics[clip=true,width=0.5\textwidth]{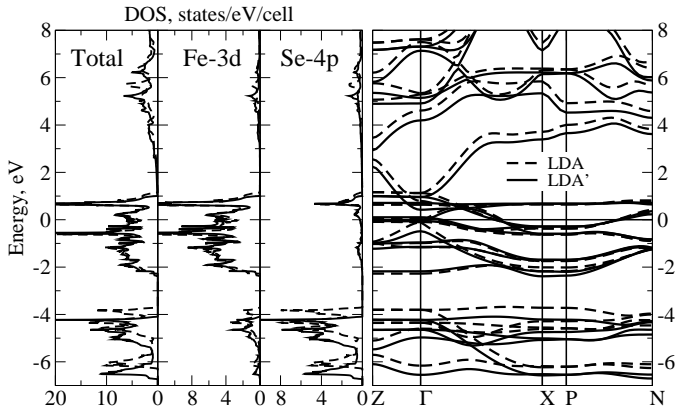}
\caption{Fig. 1. LDA (dashed lines) and LDA$^\prime$ (solid lines) calculated 
band dispersions (right panel) and total, Fe-3d and Se-4p densities of states 
(left panel) for KFe$_2$Se$_2$. The Fermi level $E_F$ is at zero energy.} 
\end{figure}

In Fig.~2 we show orbitally resolved densities of states (DOS) for Fe-3d 
orbitals. Thin gray lines represent LDA results and thin black lines -- 
LDA$^\prime$. We see that the main contribution at the Fermi level comes from 
Fe(3d)-$t_{2g}$ bands -- $xy$ and degenerate $xz,yz$
(similar to the case of Ba122 pnictide \cite{Nekr2}).
In general LDA$^\prime$ DOS'es are similar to those of LDA, except that 
LDA$^\prime$ DOS'es are few tenths of eV narrower.
In contrast to Ba122 pnictide \cite{Nekr2} both LDA and LDA$^\prime$ DOS'es
here have rather well developed ``pseudogap'' at the Fermi level.
Also at the Fermi level LDA$^\prime$ DOS is slightly higher than that of LDA.

By thick gray and black lines we show LDA+DMFT and LDA$^\prime$+DMFT DOS'es for 
corresponding Fe-3d orbitals of K$_{0.76}$Fe$_{1.72}$Se$_2$.
This hole doping level of parent compound KFe$_2$Se$_2$ corresponds to 6.08 
electrons per Fe site within LDA+DMFT calculations.
Despite correlations are moderate as compared to Fe-3d bandwidth of 
K$_{0.76}$Fe$_{1.72}$Se$_2$, we observe rather remarkable renormalization of 
the spectral weight. Interestingly, the contributions of Fe(3d)-$e_{g}$ bands 
($x^2-y^2$ and $3z^2-r^2$) in our LDA+DMFT and LDA$^\prime$+DMFT results
become larger at the Fermi level, as compared to LDA and LDA$^\prime$, while 
$t_{2g}$ bands contribution at the Fermi level remains nearly the same. 
From LDA and LDA$^\prime$ DOS'es one can note, that LDA+DMFT and 
LDA$^\prime$+DMFT results can roughly be obtained just by bands compression
around the Fermi level by a factor of 2 to 2.5, which is comparable to Ba122 
pnictide \cite{Ba122_DMFT}. However, we shall see below, that situation is not 
so simple.

\begin{figure}[!b]
\includegraphics[clip=true,width=0.45\textwidth]{KFe2Se2_comp_LDA_DMFT_DOS_comp_2.eps}
\caption{Fig. 2. Comparison of  LDA (thin gray lines), LDA$^\prime$ 
(thin black lines) and LDA+DMFT (thick gray lines), LDA$^\prime$+DMFT 
(thick black lines) densities of states for  K$_{0.76}$Fe$_{1.72}$Se$_2$ for 
different Fe-3d orbitals. The Fermi level $E_F$ is at zero energy.} 
\end{figure}

\begin{figure*}[!t]
\centering
\includegraphics[clip=true,width=0.8\textwidth]{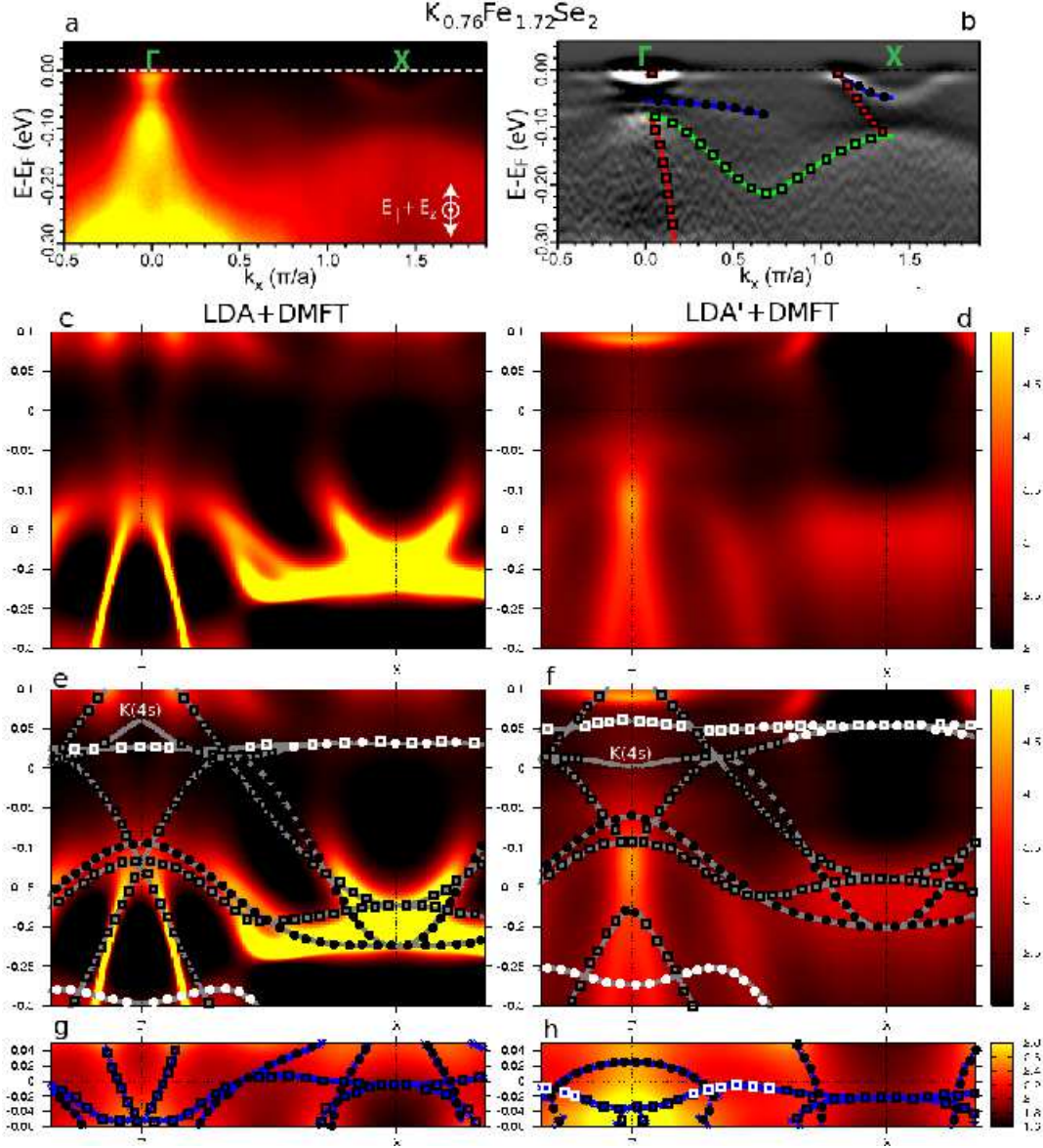}
\caption{Fig. 3. K$_{0.76}$Fe$_{1.72}$Se$_2$: comparison  LDA+DMFT (c) and 
LDA$^\prime$+DMFT (d) spectral functions with rough experimental ARPES data (a) 
and corresponding second derivative (b) from Ref.~\cite{exp}.
Panels (e) and (f) show LDA and LDA$^\prime$ bands (\modified{compressed by factor of 5}) on top of 
LDA+DMFT and LDA$^\prime$+DMFT results.
Maxima of LDA+DMFT and LDA$^\prime$+DMFT spectral functions near the Fermi level 
are presented on panels (g,h). Symbols show predominant contributions
of different Fe-3d orbitals into the spectral function (black circles - $xz,yz$,
black squares - $xy$, white circles - $3z^2-r^2$, white squares - $x^2-y^2$).
The Fermi level $E_F$ is at zero energy.} 
\end{figure*}

More detailed picture is revealed in Fig.~3. Here we show LDA+DMFT and 
LDA$^\prime$+DMFT spectral function maps for K$_{0.76}$Fe$_{1.72}$Se$_2$
along $\Gamma$-X($\pi,\pi$) high symmetry directions, compared with experimental 
data of Ref.~\cite{exp}. Both theoretical and experimental data are shown in a 
narrow energy interval from -300meV to +100meV. A common feature of theory and 
experiment is rather low intensity of the spectral function close to the Fermi 
level, where there are (almost) no well defined quasiparticle bands. It is the
main difference of K$_{0.76}$Fe$_{1.72}$Se$_2$ from similar Ba122 pnictide, 
where quasiparticle bands are clearly seen close to the Fermi level 
\cite{Ba122_DMFT}. This fact can be explained by ``pseudogap'' behavior of 
LDA+DMFT and LDA$^\prime$+DMFT DOS'es in Fig.~2 at $\pm$100meV around the Fermi 
level. This pseudogap structure is related to rather short lifetime (imaginary 
part of self-energy), together with positive inclination of the real part of 
the self-energy (Cf. Ref. \cite{pg}). This corresponds to a kind of non 
Fermi-liquid behavior close to the Fermi level.
%\modified{Corresponding values of the quasiparticle residue $Z$
%obtained at the Fermi level both on Matsubara frequencies \cite{Z} or real energies are
%of the order of 10$^{-2}$.} 

At the same time, both experiment and our theory show the pronounced 
quasiparticle bands at energies about -200meV. These bands are easily described 
by LDA or LDA$^\prime$ bands \modified{with the energy scale, which is} compressed by a factor of $\sim$5 \modified{for the energy interval $\pm$1.5~eV}. 
Corresponding results are shown on panels (e) and (f) of Fig.~3. 
Thus, K$_{0.76}$Fe$_{1.72}$Se$_2$ has much stronger quasiparticle mass
renormalization than similar 122 pnictides \cite{Ba122_DMFT}. 
We conclude, that K$_{0.76}$Fe$_{1.72}$Se$_2$ is more correlated,  than 
Fe pnictides and is rather close to Mott insulator 
(see also Refs.~\cite{DMFT_AFeSe, exp}).
To clarify LDA+DMFT and LDA$^\prime$+DMFT spectral function maps orbital 
character, on panels (e) and (f) of Fig.~3 we show with different symbols the
predominant contributions of different Fe-3d orbitals (black circles - $xz,yz$,
black squares - $xy$, white circles - $3z^2-r^2$, white squares - $x^2-y^2$).
It is also important to note, that right above the Fermi level
(at +50meV) there is rather flat  Fe(3d)-$e_{g}$ band which consists of 
$x^2-y^2$ and $3z^2-r^2$ contributions. Thus, small changes of doping level 
may lead to rather dramatic changes of electronic properties, as discussed 
previously in Refs.~\cite{Nekr_kfese,MISM}.

Now we take a closer look at the ``dark region'' of spectral density, close to
the Fermi level. If we plot the region between $\pm$50meV of the Fermi level 
on a smaller intensity scale (panels (g) and (h) of Fig.~3), we can distinguish 
some ``band'' structure, in a sense of dispersive maxima of spectral density,
which are shown in both panels. We can see here a rather narrow ``band'' of 
$xy$ symmetry. However it is not just the LDA $xy$ band, renormalized by a 
factor of 10, as proposed in Ref.~\cite{exp}. This band can be seen from second 
derivative of rough ARPES data from panel (a) of Fig.~3. Its line shape differs 
quite considerably from those of LDA. One has to remember,  that around 
the Fermi level all quasiparticle bands are rather ill defined. We also note,
that it is pretty strange that in Ref.~\cite{exp} $xz$ and $yz$ bands are found 
to be different along (0,0)-($\pi,\pi$) direction, because it is just prohibited 
by the symmetry (see also panel (a) of Fig.~3).

In Fig.~4 we present comparison of LDA+DMFT (grey lines), LDA$^\prime$+DMFT 
(black lines) Se-4p (thin lines) and Fe-3d states (thick lines) densities of 
states for  K$_{0.76}$Fe$_{1.72}$Se$_2$. The LDA+DMFT and LDA$^\prime$+DMFT 
results for the DOS are quite similar. At the Fermi level, there is a kind of 
narrow ``pseudogap'' structure for Fe-3d states. However, LDA$^\prime$+DMFT 
DOS for Se-4p states are shifted about 0.5 eV lower in energy, as compared to 
those obtained in LDA+DMFT. This suggests the importance of the measurements
of the lineshape of X-Ray photoemission spectra, as we are unaware of any 
experiments of this kind for the K$_{0.76}$Fe$_{1.72}$Se$_2$ system.

\begin{figure}[!t]
\includegraphics[clip=true,width=0.45\textwidth]{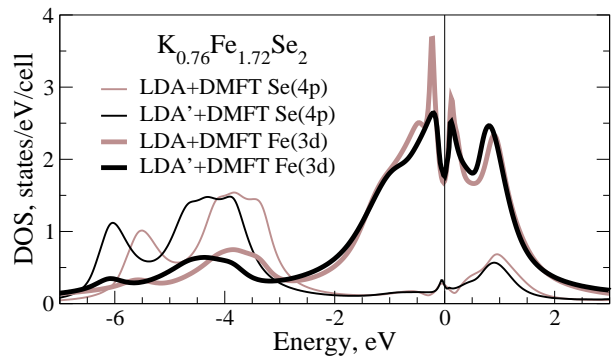}
\caption{Fig. 4. Comparison of LDA+DMFT (grey lines), 
LDA$^\prime$+DMFT (black lines) Se-4p (thin lines) and Fe-3d states 
(thick lines) densities of states for  K$_{0.76}$Fe$_{1.72}$Se$_2$.
The Fermi level $E_F$ is at zero energy.} 
\end{figure}

$Conclusion:$ In this paper we investigated the electronic structure
of hole doped high temperature iron chalcogenide superconductor
K$_{0.76}$Fe$_{1.72}$Se$_2$ in normal phase by means of LDA+DMFT and 
LDA$^\prime$+DMFT. It was found, that K$_{0.76}$Fe$_{1.72}$Se$_2$ is more 
correlated, than similar isostructural 122 Fe pnictides,  in the sense of 
bandwidth renormalization (\modified{energy scale compression by a factor about 5 in the interval $\pm$1.5~eV)
%, while at the Fermi level quasiparticle mass nearly diverges
}. Contrary to the 
conclusions of Ref.~\cite{exp}, we observe that all Fe-3d bands have the same 
bandwidth renormalization.

Also in contrast to Fe 122 pnictides, K$_{0.76}$Fe$_{1.72}$Se$_2$ compound
does not have well defined quasiparticle bands around the Fermi level
and is apparently quite close to Mott insulating phase (see also Ref.~\cite{exp}).
Both within LDA+DMFT and LDA$^\prime$+DMFT we observe a kind of ``pseudogap'' 
at the Fermi level, which is inherited from LDA band structure. The presence of 
this ``pseudogap'' leads to relatively low intensity spectral function near the 
Fermi level, which is clearly seen both in theory and in ARPES data \cite{exp}.
Similarly to Ref.~\cite{exp}, within this ``pseudogap'' region, we can 
distinguish some very low intensive ``band'' structure (both in
LDA+DMFT and LDA$^\prime$+DMFT) --- there is rather ill defined quasiparticle 
band of $xy$ symmetry, which is not simply renormalized (by a factor of 10) LDA 
$xy$ band, as proposed in Ref.~\cite{exp}. Finally we stress the importance
of measurements of photoemission spectra lineshapes.

We thank A.I. Poteryaev for providing us QMC code and many helpful discussions.
This work is partly supported by RFBR grant 11-02-00147 and was performed
within the framework of programs of fundamental research of the Russian 
Academy of Sciences (RAS) ``Quantum mesoscopic and disordered structures'' 
(12-$\Pi$-2-1002) and of the Physics Division of RAS  ``Strongly correlated 
electrons in solids and structures'' (012-T-2-1001).
NSP acknowledges the support of the Dynasty Foundation and International Center 
of Fundamental Physics in Moscow.

\end{document}